# Magnetically responsive gourd-shaped colloidal particles in cholesteric liquid crystals

Bohdan Senyuk,[a] Michael C. M. Varney,[a] Javier A. Lopez,[ab] Sijia Wang,[c] Ning Wu[c] and Ivan I. Smalyukh[*adef]

[a]Department of Physics, University of Colorado, Boulder, CO 80309, USA.
[b]Thin Film Group, University of Valle, A.A. 25360, Cali, Colombia
[c]Department of Chemical Engineering, Colorado School of Mines, Golden, CO 80401, USA
[d]Department of Electrical, Computer and Energy Engineering, University of Colorado, Boulder, CO 80309, USA
[e]Liquid Crystal Materials Research Center and Materials Science and Engineering Program, University of Colorado, Boulder, CO 80309, USA
[f]Renewable and Sustainable Energy Institute, National Renewable Energy Laboratory and University of Colorado, Boulder, CO 80309, USA

*E-mail: ivan.smalyukh@colorado.edu

## Abstract

Particle shape and medium chirality are two key features recently used to control anisotropic colloidal self-assembly and dynamics in liquid crystals. Here, we study magnetically responsive gourd-shaped colloidal particles dispersed in cholesteric liquid crystals with periodicity comparable or smaller than the particle's dimensions. Using magnetic manipulation and optical tweezers, which allow one to position colloids near the confining walls, we measured the elastic repulsive interactions of these particles with confining surfaces and found that separation-dependent particle-wall interaction force is a non-monotonic function of separation and shows oscillatory behavior. We show that gourd-shaped particles in cholesterics reside not on a single sedimentation level, but on multiple long-lived metastable levels separated by a distance comparable to cholesteric periodicity. Finally, we demonstrate three-dimensional laser tweezers assisted assembly of gourd-shaped particles taking advantage of both orientational order and twist periodicity of cholesterics, potentially allowing new forms of orientationally and positionally ordered colloidal organization in these media.

## Introduction

Colloidal particles immersed in a liquid crystal (LC) medium show a wide variety of interesting properties and phenomena.[1-50] Elastic properties of LCs trigger anisotropic interactions between colloids,[1-6] directing their self-assembly into macrostructures[7-14] useful, for example, in photonics applications.[11-13] During the last decades, experimental and theoretical studies covered different aspects of elastic interactions between spherical colloids mostly in nematic LCs.[1-15] Two-[7-11] and three-dimensional[14] crystals were designed in nematics using different elastic dipoles and quadrupoles, and singular defects were tailored using spherical[16] and topologically non-trivial[17] colloidal particles. Recent studies also showed the importance of size[18-24] and shape[24-31] in elastic interactions and ensuing assemblies of colloidal particles in nematics. On the other hand, colloidal dispersions in twisted nematic and cholesteric LCs (CLCs) have received less attention[32-44] because of the complexity of the problem caused by, for example, the screening of elastic interactions by periodic structure of cholesterics.[44] This complexity increases as the ratio between the particle's size and cholesteric periodicity increases.[32,34,39,44] Even though elastic pair interactions between colloids were thoroughly studied using optical tweezers,[5,39,42-45] elastic interactions between particles and confining walls[46-49] have rarely been explored experimentally because of a limited number of means to systematically access the exact position of the particle across the experimental sample. Moreover, the combined effects of chirality and confinement in guiding colloidal self-assembly have not been considered.

In this article, we study complex-shape colloidal particles dispersed in CLCs. These particles are gourd-shaped with dimensions larger than a cholesteric pitch, and have been functionalized to respond to magnetic fields. We experimentally probe the elastic interactions between these particles and confining walls using magnetic manipulation and elastic coupling between rotational and translational motion of these particles in CLCs. Particle–wall interactions are repulsive and generally decrease with increasing the separation between them. Intriguingly, a measured repulsive force is not monotonic with respect to increasing the separation, but exhibits an oscillatory behavior, which may be explained by screening of interactions due to the cholesteric periodicity. We show that gourd-shaped particles in CLCs reside not on a single sedimentation level, but on multiple long-lived metastable levels separated by distances comparable to the cholesteric periodicity along the helical axis. Finally, we demonstrate three-dimensional laser-tweezers-assisted assembly of gourd-shaped particles using both orientational order and the intrinsic periodicity of CLCs.

## Materials and techniques

We used gourd-shaped polystyrene nonspherical colloidal particles with two lobes of dimensions $L_1 \approx 5$ μm and $L_2 \approx 2$ μm on average (Fig. 1a–d), which were synthesized using a modified seeded polymerization technique.[51] To make them responsive to magnetic fields, they were soaked in a toluene dispersion of synthesized $CoFe_2O_4$ ferromagnetic nanoparticles[52] with a mean diameter of 10 nm for ~48 h. Ferromagnetic nanoparticles were trapped in the surface layer of gourd-shaped particles due to partial swelling of polystyrene in toluene. A cholesteric LC with a pitch (distance at which LC molecules or a director **n** twist by $2\pi$) of $p_0 = 5$ μm was prepared by mixing a nematic material ZLI-3412 and a chiral additive CB15 (both from EM Industries) in weight proportion of 96.8:3.2. The toluene was subsequently evaporated and the gourd-shaped particles then mixed into CLC and sonicated for ~5 min to break apart pre-existing aggregates. This colloidal mixture was filled in between two glass substrates spaced by glass microfibers setting the gap thickness $d = 25$ μm and $d = 60$ μm. To set planar surface boundary conditions at confining substrates we used unidirectionally rubbed thin films of spin-coated and crosslinked polyimide PI2555 (HD MicroSystem). One of the substrates was 0.15 mm thick to minimize spherical aberrations in microscopy experiments involving high numerical aperture (NA) immersion oil objectives.

We used an experimental setup built around an inverted Olympus IX81 microscope to perform bright-field, polarizing, and three-photon excitation fluorescence polarizing microscopy (3PEF-PM) observations. A tunable (680–1080 nm) Ti:sapphire oscillator (140 fs, 80 MHz, Chameleon Ultra-II, Coherent) was used for the 3PEF-PM imaging,[53,54] which in our experiments was realized by the excitation at 870 nm of CB15 molecules, which are parallel to a local **n** and whose fluorescence signal was detected within a spectral range of 387-447 nm by a photomultiplier tube H5784-20 (Hamamatsu). Problems related to beam defocusing and the Mauguin effect in the 3PEF-PM imaging[54] were mitigated by using a nematic host with low birefringence ($\Delta n = 0.078$). A scanning unit (FV300, Olympus) was used to control an in-plane position of a focused excitation beam and its polarization was varied using a half-wave plate mounted immediately before a 100× (NA = 1.4) oil objective.

Translational and rotational motion of colloidal particles was recorded with a CCD camera (Flea, PointGrey) at a rate of 15 fps and the exact $x$-$y$ position and orientation of gourd-shaped particles as a function of time was then determined from captured sequences of images using motion tracking plugins of ImageJ software. Optical manipulation and assisted assembly of gourd-shaped particles was realized with a holographic optical trapping system[53,55] operating at a wavelength of $\lambda = 1064$ nm and integrated with our optical microscopy system. Rotational manipulation of magnetically functionalized gourd-shaped colloids was achieved using an in-house custom built holonomic magnetic manipulation system integrated within the same experimental setup.[56] A magnetic field applied to a sample was measured with a LakeShore 460 3-channel gaussmeter.

## Results and discussion

### Gourd-shaped particles and magnetic orientational trapping

In the absence of external forces, gourd-shaped particles are suspended within the LC bulk and undergo thermal motion only (Fig. 1b–d). Polarizing (Fig. 1b–d) and 3PEF-PM imaging (Fig. 1f and g) show that these polystyrene colloids tend to have tangential anchoring at their surface, which dominate even after surface functionalization with ferromagnetic nanoparticles. There were as many particles oriented in the plane of the cell (Fig. 1b–d) as particles oriented at different angles out of the plane due to multiple possible metastable configurations of the director field around a gourd-shaped particles. The orientation of anisotropic gourd-shaped particles in the CLC cell (Fig. 1b–d) is determined by the local director $\mathbf{n}(\mathbf{r})$ and depends on their vertical position in the cell along a helical axis $\boldsymbol{\chi}\|z$ (Fig. 1e–i). The vertical position of the particles in the cell is dictated by the balance between forces due to elastic interactions with walls of confining substrates and the gravitational force $F_g \approx \pi(L_1^3 + L_2^3)\Delta\rho g/6 \approx 0.03$ pN, where $\Delta\rho \approx 43.5$ kg m$^{-3}$ is the difference between LC and polystyrene density, and $g \approx 9.8$ m s$^{-2}$ is the gravitational acceleration. Colloids tend to reside at the equilibrium sedimentation height in the middle of thin ($d \approx 25$ μm) cells, equally separated from both walls (Fig. 1f and g); in the thick ($d \approx 60$ μm) cells, they sink below the middle level closer to the bottom substrate (Fig. 1i). Elastic interactions with walls emerge due to significant deformations of cholesteric layers (Fig. 1f, g and i) by the gourd-shaped particles. A cholesteric layer of thickness[39,57,58] $p_0/2$ (Fig. 2a) is represented in 3PEF-PM vertical cross-section textures (Fig. 1f) as a distance between two bright or dark consecutive stripes, where the polarization of the excitation beam is, respectively, parallel or perpendicular to local $\mathbf{n}$ ($\mathbf{n} \equiv -\mathbf{n}$). Maximum displacement of the layers from their flat unperturbed state is near the particles, but decays with distance from the particle in all directions and is completely suppressed at walls with strong surface anchoring (Fig. 1f, g, i and 2a). An intrinsic pitch $p_0$ is squeezed in the deformed regions 1 and 2 to values $p_1(l,r)$ and $p_2(l,r)$ (Fig. 2), respectively, below and above a particle because a number of cholesteric layers in there remains the same as in an unperturbed region 3 far from the particle and edge dislocations were not observed (Fig. 1). Twist rates[57] $q_1(l,r) = 2\pi/p_1(l,r)$ and $q_2(l,r) = 2\pi/p_2(l,r)$ respectively in regions 1 and 2 depend on a distance from the particle $l$ and a separation $r$ between the particle and substrate. Twist rate changes nonlinearly from the smallest near the particle to the intrinsic twist rate $q_0 = 2\pi/p_0$ near the substrates (Fig. 2b). The intrinsic pitch recovers slower if the particle is further from a substrate and does not recover at all at $r \sim L_1$ (Fig. 2b).

Freely floating gourd-shaped particles exhibit both translational and rotational diffusion. Fig. 1j shows a distribution of translational displacements for an elapsed time of $\tau = 1/15$ s as measured from video tracking data collected in bright-field microscopy (Fig. 1b). Fitting the distribution by a Gaussian function[59] $P(\delta) = P_0(\delta)\exp[-\delta^2/(4D\tau)]$, where $P(\delta)$ is the probability that over the time $\tau$ a particle will displace by $\delta$, $P_0(\delta)$ is a normalization constant, and $D$ is a diffusion coefficient, allows to determine the diffusivity of gourd-shaped particles in a cholesteric material. Histograms of displacements (Fig. 1j) readily yield almost the same translational diffusion coefficients $D_x = 1.92\times10^{-3}$ $\mu m^2$ $s^{-1}$ and $D_y = 2.28\times10^{-3}$ $\mu m^2$ $s^{-1}$ in the plane of the cell and a rotational diffusion coefficient $D_\theta = 2.56\times10^{-4}$ $rad^2$ $s^{-1}$, which is coupled to the translational diffusion along the helical axis $\chi \| z$. An important feature of the system is this coupling between rotation of a gourd-shaped particle around its short axis parallel to $\chi$ and translational displacement along $\chi$, which we verified below using magnetic manipulation. This coupling allows determination of the vertical displacement of a particle along $z$-axis using the change of its orientation in the plane of the cell.

Application of a magnetic field **H** to the magnetically functionalized gourd-shaped particles results in an induced net magnetic dipole moment **m**, which allows us to use magnetic manipulation techniques[56] to control the orientation of the particles about all Euler angles (Fig. 3). When magnetically trapped, the orientation of these particles fluctuates with respect to the direction of a maximum accessible applied magnetic field equal to 10.7 mT by an angle $\Delta\theta \approx \pm 0.86°$ (Fig. 3a). The orientational trapping stiffness $k_\theta$ associated with magnetic trapping can be determined similar to the approach in optical trapping[60] using the equipartition theorem $\langle\Delta\theta^2\rangle = k_BT/k_\theta$, where $\langle\Delta\theta^2\rangle$ is a standard deviation in the orientation of a particle with respect to **H**, $k_B$ is Boltzmann's constant and $T$ is temperature. Fitting the histogram of angular deviations (Fig. 3a) with a function $f(\Delta\theta) = f_0\exp[-\Delta\theta^2/(2\langle\Delta\theta^2\rangle)]$, one finds $\langle\Delta\theta^2\rangle = 2.25\times10^{-4}$ $rad^2$, which yields $k_\theta = 18.39$ pN μm. The torque exerted by magnetic manipulation on a particle can be extracted from the time dependence of its reorientation $\phi(t)$ upon changing the direction of a magnetic field of 10.7 mT from **H**$_1$ to **H**$_2$ (Fig.

3b). The maximum angular speed of reorientation was found to be $\Omega_z \approx 0.9$ rad s$^{-1}$, so that rotational Reynolds number was small $Re_r = \rho_{LC} R_{eff}^2 |\Omega_z|/\eta \sim 10^{-8}$, where $\rho_{LC} \approx 1000$ kg m$^{-3}$ is density of LC, $R_{eff} = [(L_1^3 + L_2^3)/8]^{1/3} \approx 2.55$ μm is the radius of an equivalent sphere having the volume equal to the volume of a gourd-shaped particle, and $\eta = 146$ mPa s is the viscosity of ZLI-3412, allowing to assume non-turbulent flow and neglect inertia effects. Thus, a magnetic field torque can be balanced with a viscous drag torque as $M = (k_B T/D_\theta)\Omega_z$, which yields a dependence of torque on an angle $\Delta\phi$ between the direction of **H** and **m** (Fig. 3c). At comparably small angles, torque increases linearly with deviation of particle from trapping direction, exhibiting Hookean behavior[6] $M = k_\theta \Delta\phi$ (Fig. 3c). Using a linear fit in the Hookean regime of this dependence (Fig. 3c), one can again extract the trap stiffness $k_\theta = 13.45$ pN μm, which is in reasonable agreement with $k_\theta$ as determined above using the equipartition theorem.

The controlled rotation of gourd-shaped particle via magnetic manipulation and 3PEF-PM imaging allow for verification and determination of coupling between rotational and translational motion of gourd-shaped colloids in CLCs along $\chi \| z$ (Fig. 1k–n). The rotation of particles (Fig. 3a) was much slower than the director relaxation times.[46,48] A particle with equilibrium orientation at **H** = 0 (Fig. 1k and l) was rotated along $\chi$ counterclockwise by $2\pi$, which experimentally corresponded to a vertical displacement of the center of the particle's large lobe upwards to the top substrate along $\chi$ by a distance equal to $p_0$ (Fig. 1m and n), which yields a relation between rotational $\Delta\theta$ and translational $\Delta z$ displacements as $\Delta z = \Delta\theta q_0^{-1}$, where $q_0 > 0$ (a right-handed cholesteric) in the described experiments. One can then determine the relation between angular $\Omega_z$ and linear $V_z$ speeds as $V_z = \Omega_z q_0^{-1}$. The direction of translational displacement due to the sense of particle's rotation depends on a handedness of CLC. The right-hand rule can be applied to determine the direction of translational displacement depending on the rotation of the particle: if one grips the twist axis $\chi$ so that fingers point in the direction of particle's rotation, then, if the cholesteric helix twists with the same sense, the extended thumb points in the direction of translational displacement. Knowing the relation between rotational and translational displacements, one can determine the probability of translational displacements $\delta z$ due to Brownian motion (Fig. 1j) not only in the plane of the cell but also out of the plane along $\chi \| z$ using experimentally measured data on rotational displacements $\delta\theta$ of the gourd-shaped particles (inset of Fig. 1j). Experimental plots (Fig. 1j) and determined $D_z = 1.67 \times 10^{-4}$ μm$^2$ s$^{-1}$ show that the translational diffusion of particles in the vertical direction along $\chi$ is an order of magnitude smaller than in the plane of the cell, which was verified to be the case in both thin (Fig. 1f) and thick (Fig. 1i) LC cells.

**Particle–wall interaction**

As an important benefit of the observed coupling between rotational and translational motion of particles in our system, and because of their orientational response to magnetic field, it is possible to move particles towards either the top or bottom substrates just by rotating them via magnetic manipulation (Fig. 4). Gourd-shaped particles forced with magnetic tweezers to the bottom (Fig. 4b) or top (Fig. 4a) substrates of the thin cell ($d = 25$ μm) were repelled from the walls upon removal of the external field **H**, and subsequently moved back to their initial position in the mid-plane of the cell (Fig. 4c–g). Bright field (Fig. 4f and g), 3PEF-PM (Fig. 4a, b, d and e) microscopy and the known coupling between $\Omega_z$ and $V_z$ allow for probing a position (Fig. 4c) of particle during its motion along $\chi \| z$. Vertical displacement is not a monotonic function of time; one can clearly see a noticeable step in the plot of a position of the center of a particle's large lobe vs. time before reaching a plateau. Strong deformation of cholesteric layers near the confining walls (Fig. 4a and b) gives rise to a repulsive elastic force $F_w$ which pushes particles away from the wall into the bulk. Comparably slow motion of gourd-shaped particles across the cholesteric layers (insets of Fig. 4h and k) is overdamped and inertial forces can be neglected since a translational Reynolds number $Re = \rho_{LC} 2R_{eff}|V_z|/\eta \sim 10^{-8}$ is small. Therefore, repulsive elastic force can be balanced by a viscous drag force against the translational vertical motion of the colloidal particle as $F_w = (k_B T/D_z)|V_z|$. Effects related to hydrodynamic flow coupling to **n(r)**, which could be expected in cases of the very fast motion of a particle, can also be neglected since the Ericksen number $Er = R_{eff}|V_z|\eta/K \approx 0.016$ is much smaller than[57] 1, where $K = 12.1 \times 10^{-12}$ N is an average Frank elastic constant of ZLI-3412. Thus, using time dependence of a particle's motion from the walls (Fig. 4c) and estimated $D_z$ (Fig. 1j), one can plot the dependence of the repulsive force on a separation $r$ between the center of a gourd's large lobe and the surface. It is maximum (~8–12 pN) near the confining walls of the thin ($d = 25$ μm) cell and decreases to zero at the level where the particle reaches a sedimentation height in the middle of the cell at $z \approx 12.5$ μm (Fig. 4h and k). Integrating this force over the separation, one can determine an effective interaction potential at the onset of interactions as being in the range of tens of thousands of $k_B T$. The repulsive force does not decrease with time $t$ monotonically; there is a clearly pronounced plateau of constant force of ~ 1–2 pN along ~2 μm long path (Fig. 4h and k). The slightly stronger $F_w$ from the bottom substrate can be explained by a stronger deformation of the near surface cholesteric layer (compare Fig. 1a and b). Downward motion of the gourd-shaped particle from the top wall coincides with the direction of a gravitational force, but opposes gravity when the particle is moving upward from the bottom substrate. However, as one can see from estimated value of $F_g$, this gravitational force can be neglected in the case of thin LC cells as it is one-to-three orders of magnitude smaller as compared to the maximum elastic force (Fig. 4h and k). Magnetic manipulation is working against the repulsive force $F_w$ when moving the particle towards the wall. Thus, one can estimate an orientational trapping or escape force $F_{trap} \approx 12$ pN exerted on the particle by a magnetic orientational trap as the maximum measured repulsive force at the minimum $r$ between the particle and wall (Fig. 4h).

**Metastable periodic localization levels**

The onset of repulsive interactions between gourd-shaped particles and confining walls in thick cells (Fig. 5 and 6) is similar to that in thin cells. When released from the magnetic trap while close to substrates, colloidal particles repel quickly away from the walls (initial position 1), however, a separation between them increases non-monotonically (Fig. 5 and 6). Separation vs. time curves have a step which is even more pronounced than in thin cells; this is seen in the plots as a long plateau in the dependence up to the distance ~10-15 μm away from the walls (level 2), at which particles eventually come to the rest (Fig. 5a and 6a). What is surprising and interesting is that this metastable level is far below the equilibrium sedimentation height for particles in the thick cells (Fig. 1i). Particles can reside at this metastable level without moving up or down (depending on the bottom or top substrates) due to a local potential energy minimum arising from the interplay between gravitational and elastic forces of repulsion from both substrates and particle's interactions with the cholesteric's periodic structure. However, we find from our microscopic observations that this level is only metastable. Using the low power (<10 mW) optical trap, one can assist the particle to move slightly off this metastable level 2, so that it will start moving further towards the equilibrium sedimentation height (and never back to the substrate) by itself without further assistance of the laser tweezers, slowing down when reaching the new level 3 (Fig. 5a and 6a), which is still far away from the final equilibrium sedimentation height of ~25-30 μm. Metastable levels 2 and 3 are separated by a distance roughly equal to ~$p_0$. Plots show that the repulsive elastic force $F_w$ decreases non-monotonically with increased separation from walls (Fig. 5c and 6c). A maximum repulsive force of ~6-7 pN was acting on the particle in its location closest to the wall (level 1), similar to the thin cells, and decreased to zero when approaching the level 2. However, there was also a sudden increase of the force to ~1 pN at the separation ~11 μm (from the top substrate) (Fig. 5c) and ~7.5 μm (from the bottom substrate) (Fig. 6c) that correspond to the small steps in the time dependencies of separation (Fig. 5a and 6a) before reaching the level 2. The repulsive force increases again to ~0.5 pN during transition between levels 2 and 3 (Fig. 5c and 6c). Positions of particles on the intermediate metastable levels 2 and 3 correspond to the potential minima arising due to the particle's interaction with the helicoidal director structure of CLC mediated by the surface anchoring on particle's surface and supplemented by the competition of $F_g$ and elastic forces of repulsion from both substrates.

We fit the separation dependence of the repulsive force on the onset of elastic interactions with a power-law equation $F_w(r) \sim r^{-a}$ and find $a \approx 3$ for the repulsion from both substrates in the thin cell (Fig. 4h and k) and from the top wall in the thick cell (Fig. 5c) and $a \approx 5.2 \pm 0.18$ for the repulsion from the bottom wall in the thick cell (Fig. 6c). This difference in the short-range interactions can be explained by qualitatively different structure of cholesteric deformations in the narrow region between the particle and wall (compare Fig. 4a, b, 5b, and 6b).

Fig. 7 shows a time dependence of separation between a gourd-shaped particle and the bottom confining wall where the laser tweezers were used to assist displacing the particle away from the metastable levels until it approached the final equilibrium sedimentation level 5. There were four long-lived intermediate metastable levels, which were separated by a distance equal to $\sim p_0$, corresponding to $\sim 2\pi$ rotation of **n** (Fig. 7). Recently, similar oscillatory behavior of elastic interactions in layered systems was experimentally observed between spherical colloids in the cholesterics[44] and theoretically predicted for smectics,[50] although these interactions were of attractive type. A dimer formed from two single spheres when sedimenting in cholesterics also showed non-monotonous motion in other theoretical studies.[38] In principle, the observed periodic oscillations of repulsive elastic interactions between gourd-shaped particles and both confining walls in our system can be explained as arising due to the screening of elastic interactions by the ground-state periodic structure of CLC similar to ref. 44.

The observed behaviour is a result of a complex balance of forces acting on a gourd-shaped particle. Lavrentovich and coworkers[46,48,49] studied forces acting on a colloidal particle levitating and moving in a nematic LC confined between substrates. Colloidal particles levitate in a nematic LC at a certain distance from the bottom substrates dictated by a balance between the gravitational force and forces of elastic repulsion from bounding walls.[46,48,49] Our experimental data (see, for example, Fig. 4–7) demonstrate that a complete balance of forces acting on the gourd-shaped particle in CLC is different and more complex due to the inherent periodicity of cholesterics.

One way to interpret the experimentally observed complex elastic interactions between a gourd-shaped particle and confining wall is by considering the deformations of a periodic cholesteric structure in the regions between the particle and confining walls as well as the local within-layer **n**(**r**) deformations surrounding the particle. The cholesteric structure is significantly distorted by embedded gourd-shaped particles (Fig. 2a and 4b). Cholesteric layers are deformed from both sides of the particle and these deformations propagate towards confining walls, causing elastic interactions between the particle and both walls that tend to minimize the energetic cost of these distortions (Fig. 2). Repulsive forces acting on the particle from two walls have opposite directions. As the particle moves along the helical axis, at some particular moment, there is some configuration of cholesteric layers deformations (see, for example Fig. 6b) with corresponding $q_1(l,r)$ and $q_2(l,r)$ (Fig. 2a). It is nontrivial and beyond the scope of our present work to establish an analytical description of cholesteric layers deformations caused by an isolated inclusions on a distance from a center of the inclusion,[61,62] but we can probe it experimentally through a direct 3D imaging. Fig. 2b shows a dependence of a twist rate vs. $l$ and $r$ measured from experimental 3PEF-PM images (such as the one shown, for example, in Fig. 4a) along the line normal to cholesteric layers and going through a center of a gourd-shaped particle's large lobe. When a particle is elastically repelled from, for example, a bottom substrate (Fig. 6), a separation between them increases but $r$ between the particle and a top substrate respectively decreases. As the particle moves away from the cell wall into the LC bulk, twist rates $q_1(l,r)$ and $q_2(l,r)$ change (Fig. 2) with $r$. We observed that the twist rates change (decrease in a region 1 and increase in region 2) continuously till the moment when separation between a particle and a bottom bounding wall increases by a distance wide enough to fit another cholesteric layer (half-pitch). At the same time, the separation between a particle and a top substrate decreases so that it can fit one less of the layers (half-pitches) in-between. This change of a number of layers in regions between a particle and substrates with ensuing reconfiguration of cholesteric layers deformations and corresponding twist rates is one of the key factors leading to the oscillatory behaviour of elastic interactions observed in our experiments. Strong elastic distortions and defects immediately surrounding a particle within a cholesteric layer also contribute to the periodic potential energy landscape seen by the particle as it moves along the helical axis. The combination of the elastic forces due to these two types of distortions leads to a highly nontrivial falling motion of particles under the effects of gravity. Unlike in isotropic fluids, the velocity $V_z$ is not constant, and the particle periodically speeds up and slows down while traversing through the CLC medium due to its interaction with the periodic free energy landscape. Interestingly, the distance between metastable levels slightly decreases as the particle moves further away from the nearest confining wall, which can be seen from the orientation of the gourd-shaped particles on each such level (compare the orientation of particles in the insets 2, 3, and 4 in Fig. 7). Even though we described only a qualitative explanation of the experimentally observed and measured colloid-wall elastic interactions in CLC, a complete balance of forces acting on a gourd-shaped particle is unclear and further theoretical analytical and numerical studies are needed to characterize the precise mechanism of their oscillatory behaviour.

The distortions of cholesteric layers induced by the particles propagate to large distances (Fig. 1e–n) and thus mediate strong interactions with the confining substrates in all studied cells. Although difficult to realize experimentally, it would be of interest to explore how the particle localization behaviour could be modified or possibly disappear in a regime approaching that of an infinitely large cholesteric sample. In a confinement-free cholesteric, elastic interactions with the surfaces would not be present. Thus, dynamics of the particle moving along the helical axis could be only affected (as compared to isotropic fluids) by presence of local near-particle elastic distortions and associated surface anchoring energy costs at the particle-CLC interface. The details of this motion would depend on whether the particle would be falling and rotating by itself or along with the elastic distortions and defects that it induced, depending on material parameters of the system.

Elastic interactions between gourd-shaped particles and confining walls depend also on the orientation of particles with respect to $\chi \| z$. For example, gourd-shaped particles with their long axes tilted in the vertical plane of the cell (Fig. 8c and d) did not repel from the wall into the bulk after being moved by the magnetic trap near confining walls and subsequently released. Instead, they remained near the wall, despite the fact that a direct contact between the particle and the wall was avoided (Fig. 8d). This particle's indifference to being close to the wall can be explained by much smaller deformations of cholesteric layers, which vanish in the bulk faster before reaching the confining walls, as compared to the case when gourd-shaped particles are aligned in the plane of the cell (compare deformations of the cholesteric layers in experimental textures of Fig. 4a, d, 6b and 8d).

**Optically directed assembly of three-dimensional structures**
The observed periodicity of repulsive elastic interactions between gourd-shaped particles and confining walls and the existence of equally spaced, long-lived metastable states in CLC can be used for directed assembly of particles into complex three-dimensional colloidal structures. Advantage can be taken of not only orientation dependent elastic interactions (present also between spherical colloids in nematics),[7-14] but also taken from the intrinsic translational periodicity of cholesteric hosts. In our work, to apply these general principles, we used thick cholesteric cells, which allow for multiple metastable states below the equilibrium sedimentation height (Fig. 7). Laser tweezers were used to collect particles residing on different metastable levels (Fig. 9a, d and g) and move them adjacent to each other (Fig. 9b and c). Colloidal particles sitting on the same metastable level were oriented either in the same direction or antiparallel (Fig. 9b) showing orientational ordering defined by the orientation of **n** at the metastable level. When brought close to each other using laser tweezers, particles on different levels tended to attract to each other via elastic interactions and form colloidal structures (Fig. 9e–k) with centers of each particle's large lobe separated vertically by a distance of ~$p_0$ (difference between levels pointed by yellow arrows 1 and 2), as determined by the use of 3PEF-PM microscopy (Fig. 9h and k). Strong bending of cholesteric layers in the vertical plane (Fig. 9h and k) prevents gourd-shaped particles from coming into direct contact with each other. Blocks of three (Fig. 9e and h) and four (Fig. 9c–k) gourd-shaped particles were assembled in the vertical plane of the cholesteric cell resulting in colloidal structures that were robust and which can be translated with laser tweezers or rotated via magnetic manipulation without breaking apart.

## Conclusions

We have described the self-assembly and elastic interactions of magnetically responsive gourd-shaped colloidal particles dispersed in CLCs with a periodicity smaller than the particle's dimensions. Particles magnetically manipulated to positions near the confining walls were subsequently repelled into the bulk with a maximum elastic force of ~10 pN. We demonstrated elastic interactions of these particles with confining walls using elastic coupling between rotational and translational motion of particles in CLCs using magnetic manipulation. Particle-wall repulsive interactions were dependent on their separation, and intriguingly this measured repulsive force behaved non-monotonically with increasing separation and exhibited an oscillatory force with periodicity comparable to the intrinsic cholesteric pitch, which we explained by considering local and long-range interactions of the incorporated particle with the cholesteric periodic structure. We showed that gourd-shaped particles in CLCs can reside on multiple long-lived metastable localization levels separated by a distance comparable to the intrinsic cholesteric periodicity.

Finally, we have demonstrated three-dimensional laser tweezers assisted assembly of gourd-shaped particles using both the orientational order and translational periodicity of cholesterics, which enabled assembly of nonspherical particles in cholesteric media into low-symmetry colloidal structures with both positional and orientational ordering.

**Acknowledgements**

We acknowledge discussions with A. Mcleland, Q. Zhang, A. Martinez, Q. Liu, T. Lee, P. Ackerman, G. Zambrano, and N. P. Montenegro. This research was supported by the U.S. Department of Energy, Office of Basic Energy Sciences, Division of Materials Sciences and Engineering under Award ER46921. J.A.L. also acknowledges a partial financial support from COLCIENCIAS.

# Figures

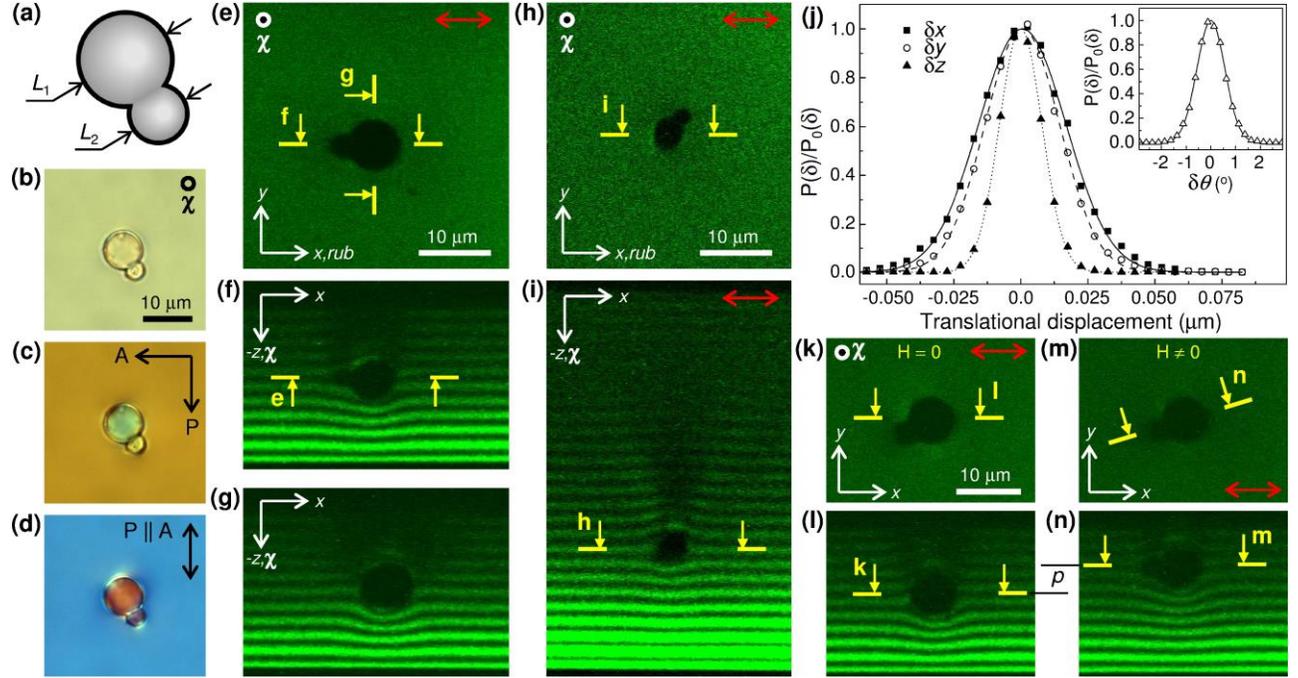

**Fig. 1.** Colloidal gourd-shaped particles in CLCs: (a) shape of particles with two lobes $L_1 \approx 5$ μm and $L_2 \approx 2$ μm; (b) bright and (c and d) polarizing microscopy textures of the particle in a cholesteric cell; (e, h, k and m) 3PEF-PM in-plane $x$-$y$ and (f, g, i, l and n) cross-sectional $x$-$z$ textures of particles, respectively, in (e–g and k–n) thin ($d \approx 25$ μm) and (h and i) thick ($d \approx 60$ μm) cholesteric cells; (j) translational diffusion properties of particles in CLC; inset shows probability of rotational displacement of a particle around a helical axis $\chi$. Displacement data were collected at a rate of 15 fps for about 10 min. Gourd-shaped particle in (m and n) is rotated using magnetic manipulation by $2\pi$ with respect to orientation in (k and l), which corresponds to a vertical displacement equal to a pitch $p_0$. A and P mark analyzer and polarizer, respectively. Red double arrow shows the direction of polarization of 3PEF-PM excitation.

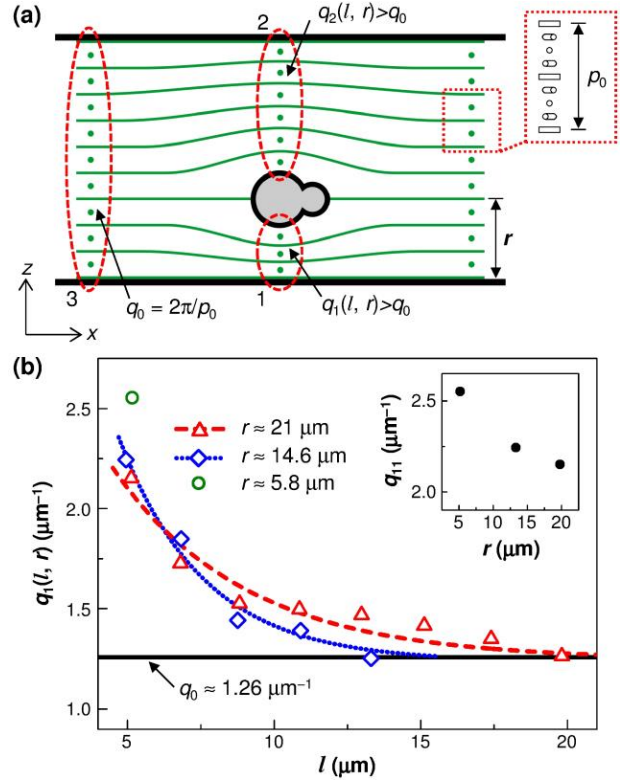

**Fig. 2.** Deformations of cholesteric around a colloidal particle: (a) schematic of a gourd-shaped particle embedded into a cholesteric cell. Two thick black lines show confining walls. Thin green lines depict the same local orientation of **n** and are separated by $p_0/2$, corresponding to cholesteric layers. Cholesteric layers are deformed in regions 1 and 2 near the particle but remain unperturbed in a region 3 far from it. The inset shows the detailed twist of LC molecules (cylinders) within $p_0$. (b) Cholesteric twist $q_1(l,r)$ depending on the distance from the particle $l$ and position in the cell $r$. The inset shows the twist rate $q_{11}$ in the nearest to the particle cholesteric layer depending on the distance r from the substrate. Red dashed and blue dotted lines are eye guides. A thick black horizontal line shows a uniform intrinsic twist across a cell far away from the particle.

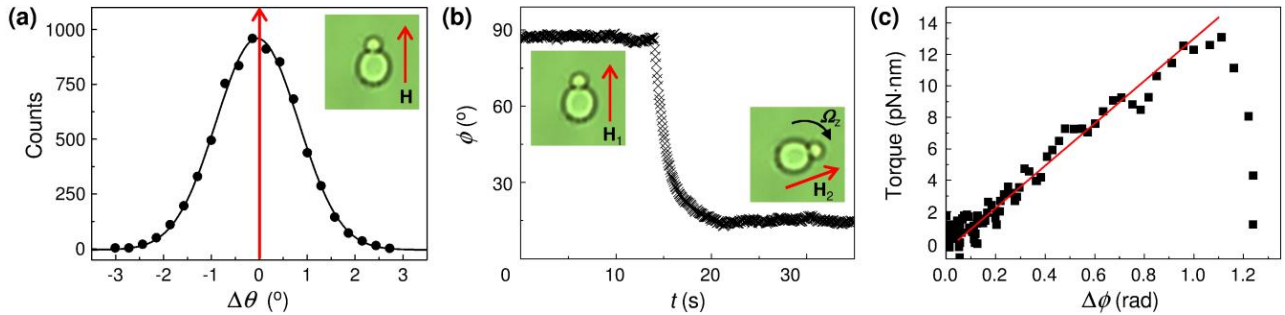

**Fig. 3.** Properties of orientational trapping of gourd-shaped particles in CLC by a magnetic field **H**: (a) histogram of angular displacements of a particle with respect to the direction (red arrow) of applied magnetic field of 10.7 mT; (b) rotation of the particle due to the change of the orientation (red arrows in insets) of trapping magnetic field from $H_1$ to $H_2$; (c) torque exerted on the particle depending on the angle between the particle's long axis and applied magnetic field of 10.7 mT as extracted from (b). Black solid line in (a) is a Gaussian fit to experimental data (black filled circles). Red solid line in (c) is a linear fit to experimental data.

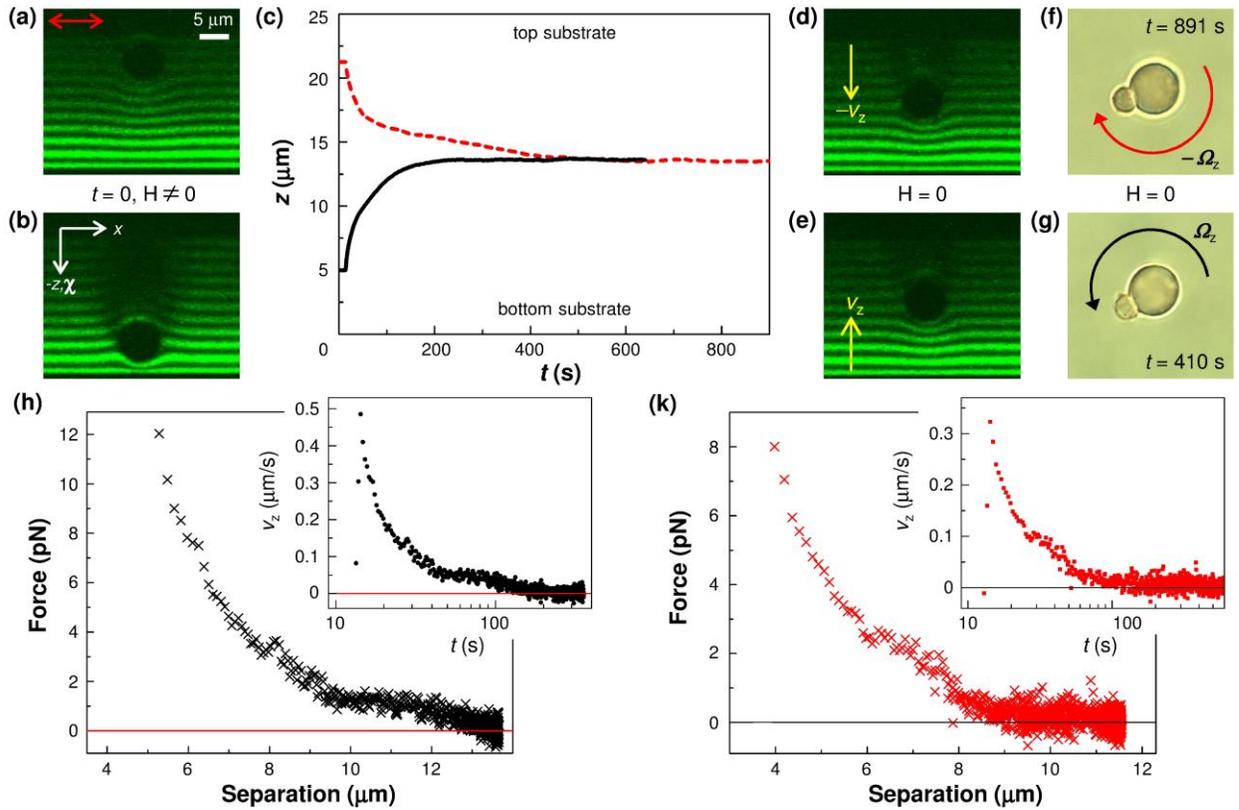

**Fig. 4.** Repulsive elastic interactions of gourd-shaped particles and confining walls in a thin ($d \approx 25$ μm) planar cholesteric cells: (a, b, d and e) 3PEF-PM cross-sectional $x$-$z$ and (f and g) bright field microscopy textures of particles in cholesteric; (c) time dependence of particle motion from walls after switching off a magnetic field **H** at $t \approx 13$ s; a repulsive elastic force from (h) bottom and (k) top walls vs. a wall–particle separation. Insets in (h) and (k) show time dependence of translational speed $V_z$ of particles.

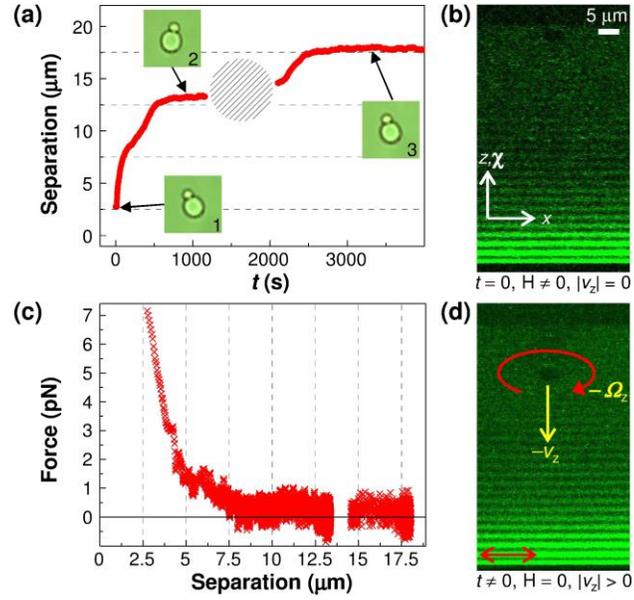

**Fig. 5.** Repulsive elastic interaction between a gourd-shaped particle and the top confining wall in a thick ($d \approx 60$ μm) planar cholesteric cell: (a) time dependence of separation from the wall after switching off a magnetic field. Bright field microscopy image 1 shows the orientation of particle at time $t = 0$; 3PEF-PM cross-sectional textures of a particle magnetically kept near a top substrate (b) and moving down (d) after switching off **H**; (c) a repulsive elastic force vs. separation from the top wall. Shaded area shows a moment when laser tweezers were applied to move the particle from a metastable level.

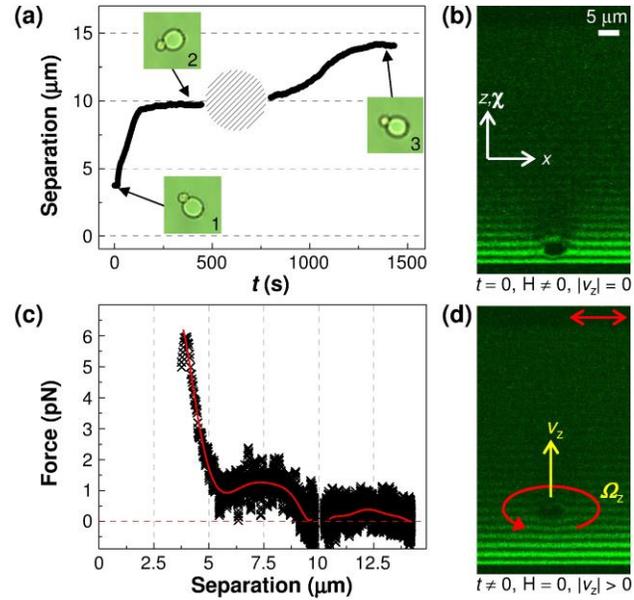

**Fig. 6.** Repulsive elastic interaction between a gourd-shaped particle and a bottom confining wall in a thick ($d \approx 60$ μm) planar cholesteric cell: (a) time dependence of separation from the bottom wall after switching off **H**; bright field microscopy images in the insets show the orientation of particle at different times indicated by arrows; 3PEF-PM cross-sectional textures of a particle magnetically trapped near a bottom substrate (b) and moving up (d) after switching off **H**; (c) a repulsive elastic force vs. separation from the bottom wall.

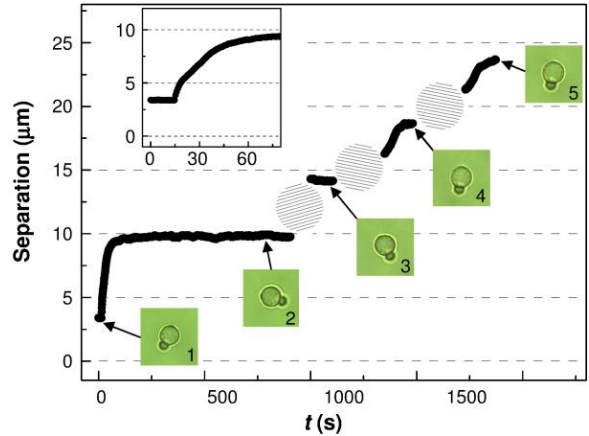

**Fig. 7.** Metastable levels of gourd-shaped colloids in a thick ($d \approx 60$ μm) planar cholesteric cell. The inset shows separation vs. time at the onset of elastic interactions. Bright field images show the orientations of particles at times indicated by arrows. Shaded areas show the moments when laser tweezers were applied to move particles from metastable states.

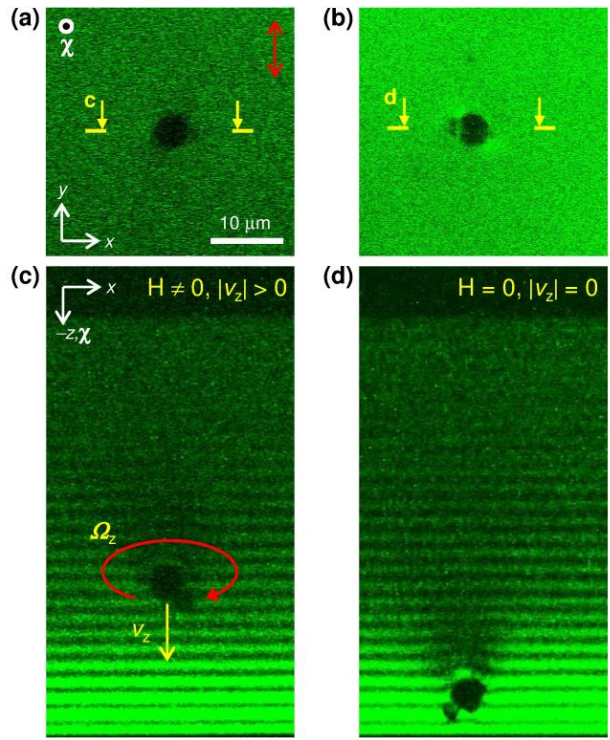

**Fig. 8.** Elastic interactions of tilted gourd-shaped particles and confining walls in planar cholesteric cells: (a and b) 3PEF-PM in-plane $x$-$y$ and (c and d) cross-sectional $x$-$z$ textures of tilted particles in a thick ($d \approx 60$ μm) cell. (c) Moving the particle down to a bottom substrate using magnetic manipulation. (d) Gourd-shaped particle staying in a metastable state near the bottom substrate after a magnetic field was switched off.

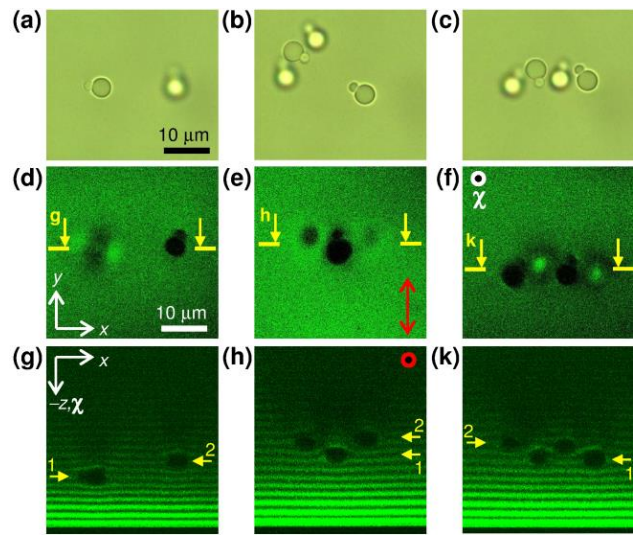

**Fig. 9.** Orientational and translational assembly of gourd-shaped colloidal particles in CLC: (a and b) bright field, (d–f) 3PEF-PM in-plane and (g–k) cross-sectional textures of assemblies formed by colloidal particles levitating at metastable levels.